\documentclass[aps,prb,twocolumn,showpacs,superscriptaddress, longbibliography, floatfix]{revtex4-2}

\usepackage{xcolor, graphicx}
\usepackage{amssymb,amsmath, amsfonts, bm}
\usepackage{newtxtext,newtxmath}
\usepackage{booktabs}
\usepackage{diagbox}
\usepackage{dcolumn}% Align table columns on decimal point
\usepackage[none]{hyphenat} % avoid the hyphen "-" that splits a word in the end of a line.
\usepackage{tabularx}

\usepackage{newtxtext,newtxmath}

\usepackage{verbatim}  % provides a comment enviroment

\begin{document}

%\title{Coexistence of quantum spin Hall states and bulk Dirac points in an inversion-asymmetric system}
\title{Large-gap quantum anomalous Hall states induced by functionalizing buckled Bi-III monolayer/Al$_{2}$O$_{3}$}

\author{Suhua Jin}
\email{These two authors contributed equally to this work.}
\affiliation{School of Physical Science and Technology, ShanghaiTech University, Shanghai 201210, China}
\author{Yunyouyou Xia}
\email{These two authors contributed equally to this work.}
\affiliation{School of Physical Science and Technology, ShanghaiTech University, Shanghai 201210, China}
\affiliation{\mbox{ShanghaiTech Laboratory for Topological Physics, ShanghaiTech University, Shanghai 201210, China}}
\author{Wujun Shi}
\affiliation{ Center for Transformative Science, ShanghaiTech University, Shanghai 201210, China}
\affiliation{Shanghai High Repetition Rate XFEL and Extreme Light Facility (SHINE), ShanghaiTech University, Shanghai 201210, China}
\author{Jiayu Hu}
\affiliation{School of Physical Science and Technology, ShanghaiTech University, Shanghai 201210, China}
\author{Ralph~Claessen}
\affiliation{Physikalisches Institut, Universit\"{a}t W\"{u}rzburg, D-97074 W\"{u}rzburg, Germany}
\affiliation{W\"{u}rzburg-Dresden Cluster of Excellence ct.qmat, Universit\"{a}t W\"{u}rzburg, D-97074 W\"{u}rzburg, Germany}
\author{Werner~Hanke}
\author{Ronny~Thomale}
\affiliation{Institut f\"{u}r Theoretische Physik und Astrophysik, Universit\"{a}t W\"{u}rzburg, D-97074 W\"{u}rzburg, Germany}
\affiliation{W\"{u}rzburg-Dresden Cluster of Excellence ct.qmat, Universit\"{a}t W\"{u}rzburg, D-97074 W\"{u}rzburg, Germany}
\author{Gang~Li}
\email{ligang@shanghaitech.edu.cn}
\affiliation{School of Physical Science and Technology, ShanghaiTech University, Shanghai 201210, China}
\affiliation{\mbox{ShanghaiTech Laboratory for Topological Physics, ShanghaiTech University, Shanghai 201210, China}}

\begin{abstract}
Chiral edge modes inherent to the topological quantum anomalous Hall (QAH) effect are a pivotal topic of contemporary condensed matter research aiming at future quantum technology and application in spintronics.
A large topological gap is vital to protecting against thermal fluctuations and thus enabling a higher operating temperature.
From first-principle calculations, we propose Al$_{2}$O$_{3}$ as an ideal substrate for atomic monolayers consisting of Bi and group-III elements, in which a large-gap quantum spin Hall effect can be realized. 
Additional half-passivation with nitrogen then suggests a topological phase transition to a large-gap QAH insulator. 
By effective tight-binding modelling, we demonstrate that Bi-III monolayer/Al$_{2}$O$_{3}$ is dominated by $p_{x}, p_{y}$ orbitals, with subdominant $p_z$ orbital contributions. 
The topological phase transition into the QAH is induced by Zeeman splitting, where the off-diagonal spin exchange does not play a significant role. 
The effective model analysis promises utility far beyond Bi-III monolayer/Al$_{2}$O$_{3}$, as it should generically apply to systems dominated by $p_{x}, p_{y}$ orbitals with a band inversion at $\Gamma$. 
\end{abstract}

\maketitle

\section{Introduction}
Since the celebrated discovery of the quantum anomalous Hall (QAH) effect~\cite{Yu61,Chang167}, the study of two-dimensional topological quantum systems and the search for new material platforms with a large topological gap has gained significant attention due to their strong potential in spintronic applications~\cite{bansil2016colloquium, RevModPhys.83.1057, RevModPhys.82.3045,RevModPhys.82.1539}.
The chiral edge state in QAH systems travels dissipationless along the edge and is believed to provide a promising solution to the Moore's Law issue in the silicon industry. 
As the analog of the integer quantum Hall effect under a strong magnetic field, the QAH acquires the quantized Hall conductivity by intrinsic magnetism without Landau Levels. 
From the theoretical proposal by F.D.M. Haldane~\cite{PhysRevLett.61.2015}, however, little progress was made in experiments on QAH systems before the field was cross-fertilized by 
the discovery of the 2D quantum spin Hall (QSH) effect~\cite{doi:10.1126/science.1133734,doi:10.1126/science.1148047} and three-dimensional (3D) topological insulators (TIs) ~\cite{RevModPhys.82.3045,zhang2009topological,doi:10.1126/science.1173034,PhysRevB.81.205407,zhang2010crossover}. Note that the QAH is distinctly different from Chern insulator realizations in classical platforms, where the dissipationless character of the topological edge modes is not found~\cite{marin}.
The discovery of the QSH and 3D TIs implicitly paved the way for the realization of the QAH in experiments, as the field turned its focus to materials with significant spin-orbit coupling and a careful analysis of bulk and surface effect in semiconductor composites~\cite{Chang167, checkelsky2014trajectory,PhysRevLett.113.137201,chang2015high,PhysRevLett.118.246801,PhysRevB.103.235111}. 

QSH systems and 3D TIs are time-reversal (TR) invariant topological systems. Their Hamiltonian is block-diagonal with different angular momenta $j=l\pm1/2$ being TR partners. 
They form two subsystems that are also individually topological. 
Each subsystem can be viewed as a QAH system. Their combination under TR symmetry generates the QSH or the surface states in 3D TIs. 
Thus, breaking TR can potentially revert the above combination to yield the QAH. 
The essential condition of this process is to retain one of the subsystem's topological characters while dissolving the other one~\cite{Wang:2015wr, Liu:2016tz, KOU201534, PhysRevB.78.195424}.
Theoretical proposals on magnetically doped HgTe/CdTe ~\cite{PhysRevB.85.125401} and InAs/GaSb ~\cite{PhysRevLett.113.147201} quantum wells nicely fulfilled the required criteria.
Unfortunately, experimental attempts did not succeed in realizing the QAH due to the failure in establishing long-range ferromagnetic order. Summarizing this direction of attempts, the immediate manipulation of the 2D QSH states does not seem particularly promising in achieving QAH at first sight. 
Instead, the final breakthrough was made along the other route, i.e., magnetically doping the surface states of 3D TIs~\cite{Yu61, Chang167}.
Experiments found that doping tetradymite semiconductors Bi$_{2}$Te$_{3}$, Bi$_{2}$Se$_{3}$, and Sb$_{2}$Te$_{3}$ with Cr, Mn, and V can significantly enhance the bulk spin susceptibility and establish a ferromagnetic insulating phase through the van Vleck mechanism~\cite{Yu61, Chang167, PhysRevB.81.195203, KOU2013, KOU2013-2, PhysRevLett.114.146802, PhysRevB.71.115214}. 
At the same time, the ferromagnetic exchange coupling makes one copy of the QAH subsystem revert to normal band order and, therefore, lose its nontrivial topology. Consequently, only one edge state survives, leading to the QAH. 
In recent years, the MnBi$_{2}$Te$_{4}$-family materials have also provided a platform for QAH with intrinsic magnetism from the inserted MnTe bilayer~\cite{Luo2013FromOL}.
The interplay between the magnetic coupling and the topologically nontrivial bands endows the MnBi$_{2}$Te$_{4}$-family materials with rich topological phases~\cite{PhysRevLett.122.107202, PhysRevLett.122.206401, doi:10.1126/sciadv.aaw5685,10.1093/nsr/nwaa089, PhysRevX.9.041038, PhysRevLett.124.167204, PhysRevLett.124.197201, PhysRevLett.124.126402, PhysRevX.10.031013, PhysRevLett.126.176403, PhysRevLett.123.096401, PhysRevX.11.011039}.

Despite the successful realization of the QAH in magnetically-doped tetradymite semiconductors and MnBi2Te4-family materials, the operating temperature is limited to a relatively low value. Therefore, new QAH materials or QSH parent compounds with large topological gaps are strongly sought after.   
A large gap QSH state was experimentally realized in the monolayer bismuthene system~\cite{Reis287, PhysRevB.98.165146}. 
This system is geometrically equivalent to graphene, while a completely different mechanism characterizes its low-energy excitation.
The $\sigma$-bond states form an effectively two-orbital Kane-Mele model, which encourages a large onsite spin-orbital coupling (SOC) that supports a topological gap as large as 0.8 eV. 
 Bismuthene is a QSH system with the so-far largest topological gap confirmed in the experiment, which thus can potentially be used in room-temperature spintronic device applications. 

\begin{figure*}[t]
\centering
\includegraphics[width=\linewidth]{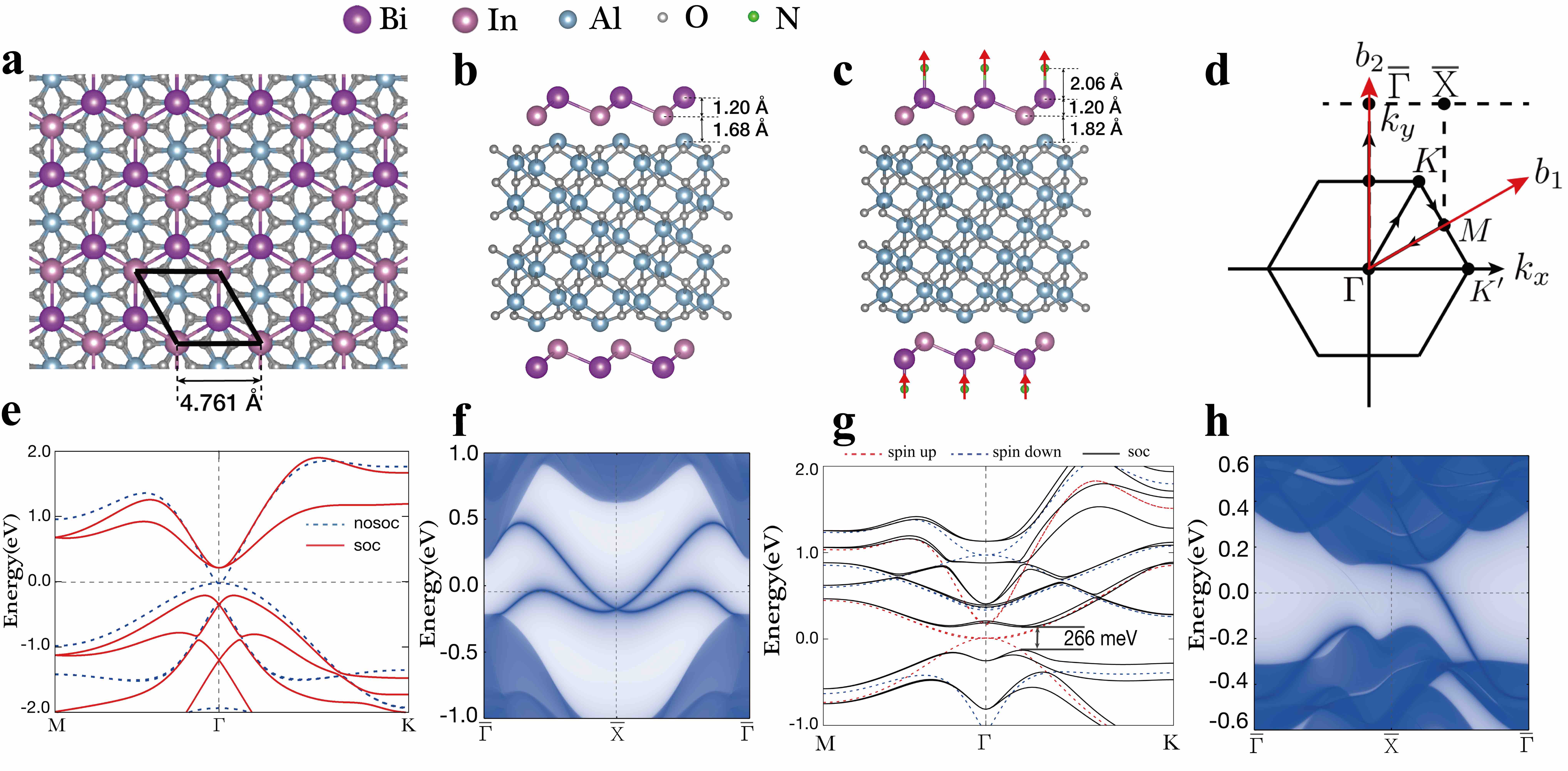}
\caption{(Color online) (a-b) Structural model of BiIn/Al$_2$O$_3$ in top and side views. (c) Half-passivation of BiIn with nitrogen. (d) The 2D Brillouin zone and the high-symmetry k-path adopted in the calculation. (e) Electronic structure of BiIn/Al$_{2}$O$_{3}$ without SOC (blue dashed line) and with SOC (red solid line). (f) Calculations of topological edge mode of a ribbon with zigzag termination. (g) and (h) are same as (e) and (f), but for nitrogen half-passivated BiIn/Al$_{2}$O$_{3}$.}
\label{Fig1}
\end{figure*}

This paper would like to revisit 2D QSH parent compounds as a hitherto less successful recipe to reach the QAH. However, as a game-changing starting point, we employ the room-temperature QSH in bismuthene as the parent state for our QAH design.
We start from a large-gap QSH system and introduce the long-range ferromagnetism by breaking the complete shell configuration of certain atoms, such that TR will be broken. 
Our idea is inspired by recent studies on several 2D materials with honeycomb structure of elements of groups IV~\cite{PhysRevLett.111.136804,Chou_2014,wang2016tunable}, V~\cite{PhysRevB.90.085431,song2014quantum,jin2015quantum,C4CP02213K,hsu2016two}, and III-V~\cite{chuang2014prediction,crisostomo2015robust,yao2015predicted,li2015giant,zhao2015driving,PhysRevB.91.235306, PhysRevB.91.165430, PhysRevB.91.165430,jin2015quantum}.
QAH states were found in buckled Bi-III honeycomb systems by chemically decorating one side with hydrogen and the other side with nitrogen.  
As a result, a net magnetic moment can be established~\cite{crisostomo2017chemically}.
Furthermore, we employ the fact that functionalizing semiconductors with nitrogen is an experimentally feasible operation; for example, it induces a magnetic moment in 3D TI Bi$_2$Te$_3$~\cite{Jin_2012}. 
Thus, introducing charge imbalance and breaking complete shell configurations by functionalizing large-gap QSH insulators with nitrogen atoms may result in a large-gap QAH scenario.

We propose the experimentally feasible monolayer/substrate combination, i.e., Bi-III monolayer on Al$_{2}$O$_{3}$, as a new platform for realizing a large-gap QSH.
It represents a promising material candidate similar to recent proposals~\cite{xia2021hightemperature}, with high experimental feasibility for epitaxial synthesis. 
We further tune these QSH systems to large-gap QAH insulators by functionalizing them with nitrogen with the largest bandgap as large as 405 meV. 
In particular, we provide a unified, effective Hamiltonian description of QSH and QAH states that originate from 2D honeycomb systems with low-energy excitations dominated by $\sigma$-bonds. 
%Thus, the essence of our work is twofold: One is the proposal of Al$_{2}$O$_{3}$ as an ideal substrate for growing Bi-III monolayer. The other one is the presentation of a generic Hamiltonian for this class of materials. 

Our paper is organized as follows.
In Section~\ref{Sec:QSH}, we propose $\alpha$-Al$_{2}$O$_{3}$ as the appropriate substrate for Bi-III monolayer that a large-gap QSH can be realized. 
Based on these systems, we further modify them by functionalizing bismuth with nitrogen to introduce long-range ferromagnetism.
In Section~\ref{Sec:model}, we provide a detailed understanding of the QSH and QAH in Bi-III monolayer/Al$_{2}$O$_{3}$, as well as the topological transition between them by explicitly constructing an effective tight-binding model.

\section{Material platforms}
\label{Sec:QSH}
Corundum, i.e. $\alpha-$Al$_2$O$_3$, is a typical substrate material, and $\alpha$ is its most stable phase in nature. 
It is widely used in experiments to epitaxial grow topological thin films~\cite{WOS:000375889700047, WOS:000209956000002, WOS:000335720300038, WOS:000501493600006}.
The experimental lattice constant of Al$_{2}$O$_{3}$ is $4.761$Å~\cite{peintinger2014quantum}. 
The calculated equilibrium lattice constant of the three Bi-III monolayer systems studied in this work are 4.928 Å (BiTl), 4.805 Å (BiIn), and 4.521 Å (BiGa)~\cite{chuang2014prediction}. 
Thus, the applied strains on Bi-III monolayers range from $-5\%$ to $3.5\%$, indicating high feasibility for experiment to grow them on Al$_{2}$O$_{3}$. 

Our first-principles calculations are based on the generalized-gradient approximation (GGA) in the Perdew-Burke-Ernzerhof (PBE) form~\cite{PhysRev.136.B864, PhysRev.140.A1133, PhysRevLett.45.566, PhysRevB.23.5048, PhysRevLett.77.3865} within the framework of the density-functional theory (DFT) using projector-augmented-wave (PAW)~\cite{PhysRevB.59.1758} wave functions as implemented in the Vienna Ab-Initio Simulation Package (VASP)~\cite{PhysRevB.47.558,PhysRevB.54.11169}. 
The effect of Van-der-Waals (VdW) interactions was taken into account by using the empirical correction scheme of Grimme (DFT-D2)~\cite{https://doi.org/10.1002/jcc.20495}. 
The cut-off energy for the wave functions was set at 500 eV. To simulate the experimental situation, we only allow the atomic positions of atom Bi and group III element to relax while keeping the lattice constant and the atomic positions of the substrate unchanged. Atomic positions were optimized for each lattice constant value considered until the residual forces were less than $5 \times 10^{-3}$ eV/Å. The self-consistent convergence threshold for total energy was set at $10^{ -5 } $ eV. A vacuum layer of at least 20 Å along the z-direction was used to simulate thin films. A $\Gamma$-centered Monkhorst-Pack~\cite{PhysRevB.13.5188} grid of $9 \times 9 \times 1$ was used for 2D integrations in the Brillouin zone (BZ). In the phonon calculation, we use $5 \times 5 \times 1$ supercell, and the self-consistent convergence threshold for total energy was set at $10^{ -7 } $ eV. 
Finally, the edge states were calculated by using our in-house code {\it LTM} (\mbox{\underline{L}ibrary for \underline{T}opological \underline{M}aterial calculations}) with the iterative Green's function approach~\cite{Sancho_1985} based on the maximally localized Wannier functions~\cite{PhysRevB.56.12847} obtained through the VASP2WANNIER90~\cite{MOSTOFI2008685}.

\subsection{QSH}
As Bi-III buckled monolayers alone are QSH systems~\cite{chuang2014prediction}, an appropriate substrate should less correlate with the Bi-III monolayers to maximally keep the topology of the latter. 
We, thus, consider the Al-terminated (0001) surface of Al$_{2}$O$_{3}$ as observed in experiments in our present work. 
If an O-terminated surface is used, additional hydrogen passivation would be needed~\cite{PhysRevB.84.155406}. 
All three Bi-III/Al$_2$O$_3$ systems studied in our work share similar conclusions. Therefore, we only use BiIn/Al$_2$O$_3$ as an example in the main text. 
Results on other systems can be found in the Supplementary Information. 

Fig.~\ref{Fig1}(a) and ~\ref{Fig1}(b) illustrate the crystal structure of BiIn/Al$_2$O$_3$. 
Here, we adopt a symmetric structure to simulate a semi-infinite substrate. Otherwise, the other side of the substrate needs to be carefully saturated to avoid any dangling state to interfere with the Fermi level.  
The BiIn monolayer binds to the Al-terminated surface of Al$_{2}$O$_{3}$ via VdW force. 
Full relaxation of the BiIn layer results in a separation of 1.68 \AA ~from the substrate surface and a bucking of 1.20 \AA~amplitude between the Bi and In layers. 

The corresponding bulk electronic structure is displayed in Fig.~\ref{Fig1}(e) with the BZ shown in Fig.~\ref{Fig1}(d).
In the energy range shown in Fig.~\ref{Fig1}(e), all states are from BiIn. The substrate states are far away from the Fermi level without interfering with the topology of BiIn monolayer. 
Without SOC, the system is a semi-metal with two parabolic bands touching at the $\Gamma$ point. 
When SOC is included, the system becomes a semi-conductor and opens a bandgap as large as 420 meV, which is smaller than bismuthene/SiC(0001)~\cite{Reis287}, but much larger than HgTe ~\cite{doi:10.1126/science.1133734,doi:10.1126/science.1148047} and InSb/GaAs ~\cite{knez2012quantum}. 
This gap is topological nontrivial. 
A direct calculation of a semi-infinite ribbon with a zigzag boundary gives two counter-propagating edge modes connecting the valence and conduction bands, which is the characteristic of 2D QSH insulators. 
The topological edge states reside right in the middle of the bulk bandgap, which is of high utility for device application.  

In contrast to bismuthene/SiC(0001), the nontrivial topology here stems from the band inversion of $p_{x/y}$ orbitals.  
In supplementary Fig.~S4, we show the orbital components for the states around the Fermi level, which are governed by all three $p$ orbitals of Bi and In. 
Thus, the low-energy model description of BiIn/Al$_{2}$O$_{3}$ will not be the simple $p_{x}$ and $p_{y}$ model~\cite{Reis287, PhysRevB.98.165146}.
Due to the buckling structure of the BiIn monolayer, $p_{z}$ enters as an active orbital. 
However, the band inversion is between $p_{x/y}$. Thus, the $p_{z}$ orbital plays a marginal role here.
Nevertheless, due to the participation of the $p_{z}$ orbital, the topological gap is no longer proportional to the onsite strength of $l_{z}s_{z}$ SOC. 
Thus, the size of topological gaps is reduced in these buckled system as compared to bismuthene/SiC(0001)~\cite{Reis287}. 

\subsection{QAH from chemical functionalization}
There are two strategies to break TR symmetry and introduce QAH in two-dimensional systems, i.e., by passivation with functionality and magnetic doping with transition-metal atoms. 
So far, the candidate QAH materials theoretically proposed include HgTe quantum well magnetically doped with Mn~\cite{PhysRevLett.101.146802, PhysRevB.88.085315}, (Bi, Sb)$_{2}$Te$_{3}$ doped with Cr and V~\cite{Yu61}, group-IV and V honeycomb monolayers passivated by transition atoms~\cite{PhysRevB.82.161414, PhysRevB.89.035409, Kaloni2014, jin2015quantum, PhysRevLett.113.256401, Chen2016, Huang2018, HUANG2020246} or proximity to magnetic substrates~\cite{ PhysRevLett.112.116404} and the intrinsic magnetic intercalation in MnBi$_{2}$Te$_{4}$ family materials~\cite{PhysRevLett.122.107202, PhysRevLett.122.206401, doi:10.1126/sciadv.aaw5685,10.1093/nsr/nwaa089, PhysRevX.9.041038, PhysRevLett.124.167204, PhysRevLett.124.197201, PhysRevLett.124.126402, PhysRevX.10.031013, PhysRevLett.126.176403, PhysRevLett.123.096401, PhysRevX.11.011039}.. 
Some of them have been experimentally examined and confirmed~\cite{Chang167, chang2015high, checkelsky2014trajectory, PhysRevLett.113.137201}.
Their success stimulates our search for the QAH candidate systems with a larger topological gap and higher operating temperatures. 
After obtaining the large-gap QSH, we further explored the half-passivation of BiIn monolayer with H, OH, F, Cl, Br, I, and N to search for the QAH. 
We find that only N-adsorption yields a substantial charge transfer to nitrogen atoms from bismuth and indium atoms. A charge imbalance between the spin-up and spin-down electrons on nitrogen forms which breaks the TR symmetry. 
Electrons at different nitrogen sites do not hop directly but through indium and bismuth atoms. 
Thus, the ferromagnetic couplings are mediated by the superexchange mechanism and favored by the Goodenough-Kanamori rule~\cite{PhysRev.100.564, KANAMORI195987}. 

As illustrated in Fig.~\ref{Fig1}(c), the nitrogen atom connects to bismuth but not to indium, i.e., half-passivation. 
To remove the influence of electric dipole moment, we took a structure of Fig.~\ref{Fig1}(c) with space inversion symmetry by sandwiching Al$_2$O$_3$ with N-Bi-III on both sides. 
Spin-polarized band calculation without SOC shows that chemical adsorption of nitrogen leads to an intrinsic magnetic moment of 2.0 $\mu_B$ per unit cell due to the uneven distribution of the additional charge between the spin-up and spin-down states of the nitrogen atom. 
The ferromagnetic state is energetically more stable than the nonmagnetic and antiferromagnetic states.
Local moments mainly form at N and In. 
The Curie temperature is theoretically estimated as $T_{c}\sim 110~K$ with classical Heisenberg model.
As a consequence of the ferromagnetism, we obtain a half-metal, i.e., spin-up bands are gapless while spin-down bands are gapped. 
Two spin-up bands degenerate at the $\Gamma$ point. 

\begin{figure}[t]
\centering
\includegraphics[width=\linewidth]{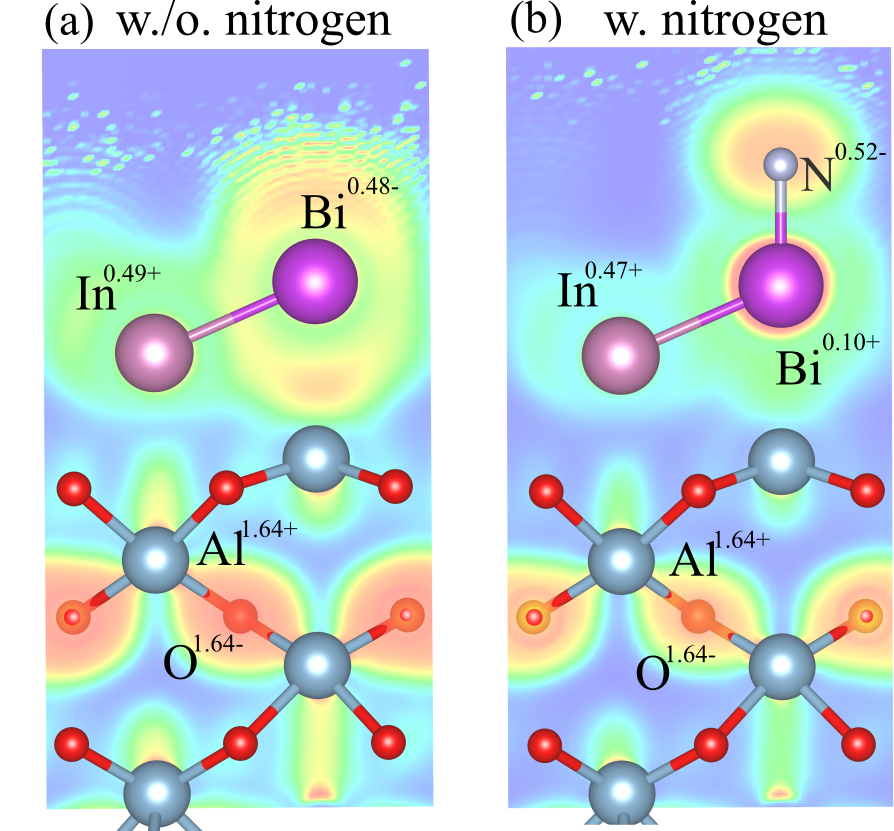}
\caption{Electron localization function (ELF) of BiIn/Al$_{2}$O$_{3}$ before and after nitrogen adsorption. The crystal model overlays on the ELF map showing the valence charge centers. Al$_{2}$O$_{3}$ substrate displays the expected ionic-bonding nature with the positive and negative valences at Al and O sites, respectively. Bi-In and In-N bonds share valence electrons displaying a strong covalent bond nature. The signed number on the shoulder of each element denotes its valence.}
\label{Fig_ELF}
\end{figure}

Before discussing the topological nature, we first explain how the local magnetic moments form due to the half-passivation of nitrogen.
To this end, we calculated the electron localization function and the ionic charge of Bi, In, and N from the Bader-Charge method~\cite{doi:10.1063/1.3553716,HENKELMAN2006354,sanville2007improved,Tang_2009}.
The former helps to understand the bonding nature of Bi, In, and N. The latter directly tells us the change of valence charge after nitrogen adsorption.  
Without nitrogen, bismuth/indium gains/loses valence electrons, respectively, forming valence Bi$^{0.48-}$ and In$^{0.49+}$. Such valence configurations are far from their formal valence. Thus, they will not form an ionic bond. 
As shown in Fig.~\ref{Fig_ELF} (a), bismuth and indium share valence charges and form a covalent bond with strong metallic nature in between.  
The Al-O bond in the substrate, however, remains ionic, and BiIn monolayer bonds with Al$_{2}$O$_{3}$ via vdw bonding.
After adsorption with nitrogen, the valence of indium does not change, but bismuth transfers valence electrons to nitrogen.
Similar to the Bi-In covalent bond shown in Fig.~\ref{Fig_ELF} (2), after nitrogen adsorption Bi-In bond remains covalent with metallic nature. 
They form a positive valence center bonding with the negative valence center N$^{0.52-}$, indicating a strong hybridization of the electronic states on these atoms. 
Thus, the charge polarization at the nitrogen site (see Tab.~\ref{Table-1}) will lift the spin degeneracy of the electronic structures.
As a result, bismuth and indium stay in the intrinsic magnetic field created by nitrogen charge polarization, and the electronic structure becomes spin-polarized.
The strong hybridization of the nitrogen, bismuth, and indium orbitals can also be seen from the projective electronic structure in Fig.~S3 of the Supplementary Information.

We also find that, except for the different binding energy levels, the spin-up and spin-down band curvatures are slightly different, implying a weak spin-exchange interaction between the spin-up and spin-down electrons. 
Only Zeeman splitting plays a dominant role. 
As BiIn is a QSH insulator, both spin-up and spin-down bands are topologically nontrivial. 
Each contributes a dissipationless edge mode. The two edge currents propagate in opposite directions and relate by the TR symmetry. 
Changing such a QSH state to a QAH state is sufficient to shift one spin component away from the Fermi level. 
Thus, there is no need for exchange coupling to transform topologically trivial bands into nontrivial ones.
%With SOC, BiIn/Al$_{2}$O$_{3}$ exhibits a bandgap of a considerably large size. 
%Among the three material systems studied in this work, the N-BiGa system possesses the largest bulk gap of 405 meV (See supplementary Fig.~\ref{FigS2}).
%The atomic SOC strength of Tl and In are stronger than that of Ga, while the band gaps of N-BiTl and N-BiIn are smaller. 
%This is because the bands contributed by $p_z$ orbitals of BiTl and BiIn appear in the bandgap formed by %$p_x$ and $p_y$ orbitals, which reduces their global bandgap.

\begin{table}
\centering
\renewcommand\arraystretch{1.7}
\renewcommand\tabcolsep{5.0pt}
\caption{The gain/loss of valence electrons (-/+) before and after nitrogen-adsorption, the charge polarization, and the magnetic moments of N, Bi and the III group elements. }
\label{Table-1}

\begin{tabular}{cccccc} \toprule
           &    & \multicolumn{2}{c}{Ionic charge change} & Charge Pola. & M ($\mu_B$)  \\ \midrule
         & N  &\diagbox {}{}  & 0.57-  & +1.07  &  0.822     \\
N-BiTl   & Bi & 0.20- & 0.25+    & -0.14 &  -0.012     \\
         & Tl & 0.22+ & 0.38+   & +0.07 & -0.003       \\
         \hline
         & N  & \diagbox {}{}  & 0.52- & +1.11   &  0.877     \\
N-BiIn   & Bi & 0.48- & 0.10+  & -0.16  & -0.021     \\
         & In & 0.49+ & 0.47+  & -0.03  & -0.039     \\
         \hline
         & N  & \diagbox {}{}  & 0.56-  & +1.08  &  0.827     \\
N-BiGa   & Bi & 0.42- & 0.18+  & -0.12  & -0.011     \\
         & Ga & 0.49+ & 0.41+  & +0.05  & -0.005     \\
\bottomrule
\end{tabular}
\end{table}

After understanding the origin of the ferromagnetic long-range order and the fully spin-polarized low-energy excitations,  we further calculated the edge states with a zigzag termination in Fig.~\ref{Fig1}(h). A helical edge state exists connecting the valence band and conduction band corresponding to a Chern number $C=1$. 
More importantly, for the BiIn topological bandgap shown in Fig.~\ref{Fig1}(g) remains as large as 266 meV, which still promises an excellent chance for the observation of the QAH in experiments.
We note that nitrogen doping/adsorption is feasible in experiments and has successfully induced ferromagnetic order in various semiconductors~\cite{doi:10.1021/jacs.6b12934, doi:10.1021/jp303465u, Liu2013, zhang2018nitrogen}. 
The QAH states in nitrogen adsorbed BiIn/Al$_{2}$O$_{3}$ benefit from two features of the N-functionalization. 
 First, the nitrogen spin-polarization breaks the TR symmetry; Second, it does not induce additional states at the Fermi level. Consequently, a topological gap formed by spin-polarized band structure is obtained, which leads to the single helical edge mode.
%Our first-principle material calculations demonstrate a topological phase transition from a large-gap QSH state to a large-gap QAH state. 
%The presence of a sizeable topological gap is similar to what we observed in bismuthene~\cite{Reis287, PhysRevB.98.165146}, 
%however, the mechanism for the emerging magnetic moment and the topological phase transition have still to be clarified. 
%Especially for the latter, the mechanism driving the two counter-propagating edge modes to just one chiral method remains to be understood. 
%In the following, we explain the above questions analytically by constructing a low-energy tight-binding model.

\section{Tight-binding Model}
\label{Sec:model}

To better understand the low-energy physics and the topological transition between the QSH and QAH in N-BiIn monolayer/Al$_2$O$_3$, we analytically construct a tight-binding model (TBM) based on all p-orbitals of Bi and In, which reproduces the DFT electronic structure and the topological nature of the above DFT calculations.    
We took the basis functions
$\left|p_{x \sigma}^{Bi}\right\rangle,\left|p_{y \sigma}^{Bi}\right\rangle,\left|p_{z \sigma}^{Bi}\right\rangle,\left|p_{x \sigma}^{In}\right\rangle,\left|p_{y \sigma}^{In}\right\rangle,\left|p_{z \sigma}^{In}\right\rangle$, 
and construct an effective Hamiltonian with the following form:
\begin{eqnarray}\label{Eq:model}
\mathcal{H} & =& H_{0} + H_{M} + H_{SO}\;, \nonumber\\
H_{0}&=&\sum_{i\alpha\sigma} \varepsilon_{i\sigma}^{\alpha} c_{i\sigma}^{\alpha, \dagger} c_{i\sigma}^{\alpha}+\sum_{\langle i, j\rangle\alpha\beta\sigma} t_{i j \sigma}^{\alpha\beta}\left(c_{i\sigma}^{\alpha,\dagger}c_{j\sigma}^{\beta}+h . c .\right)\;, \nonumber\\
H_{M}&=&- \sum_{i\sigma\sigma^{\prime}} \lambda_{m}^{i} c_{i \sigma}^{\dagger} c_{i \sigma^{\prime}}[\hat{\mathbf{m}} \cdot \hat{\mathbf{s}}]_{\sigma \sigma^{\prime}}\;, \nonumber\\
H_{SO}&=& \sum_{i\alpha\beta\sigma\sigma^{\prime}} \left\langle \alpha \sigma| \lambda_{SO} \vec{L} \cdot \vec{S} |\beta \sigma^{\prime}\right\rangle c_{i\sigma}^{\alpha, \dagger}c_{i\sigma^{\prime}}^{ \beta}\;. 
\end{eqnarray}
$H_{0}$ denotes the tight-binding model resulting only from single-particle hoppings.
$\varepsilon_{i\sigma}^{\alpha}$, $c_{i\sigma}^{\alpha}$/$c_{i\sigma}^{\alpha, \dagger}$ represent on-site energy, electron annihilation/creation operators at the $\alpha$-orbital and the $i$th atom, respectively. 
$t_{i j \sigma}^{\alpha \beta}$ is the coupling strength of electrons with spin $\sigma$ at the $\alpha$-orbital of the $i$th atom with those electrons at the $\beta$-orbital of the $j$th atom, which can be easily calculated from Slater-Koster integrals~\cite{PhysRev.94.1498}.
$H_{M}$ and $H_{SOC}$ are the corresponding magnetic coupling and SOC terms. 
$\vec{L}$ and $\vec{S}$ denote the orbital and spin angular momentum operators, and $\lambda_{SO}$ is the strength of SOC. 
The potential difference between A and B sublattices reflects the corresponding environmental difference around Bi and In, breaking sublattice symmetry.
We note that $H_{0}$ and $H_{SOC}$ are similar to the tight-binding model of Bi-III monolayer on SiO$_{2}$~\cite{xia2021hightemperature}.
Additionally, in the current model, a TR symmetry breaking term $H_{M}$ is also introduced, accounting for the Zeeman splitting, which shifts the two spin-degenerate bands up and down with an equal amount of energy.
$\hat{\mathbf{m}}$ is the direction of magnetic polarization and $\lambda_{m}^{i}$ is the magnitude of magnetic polarization.
 
 \begin{figure}[htbp]
\centering
\includegraphics[width=\linewidth]{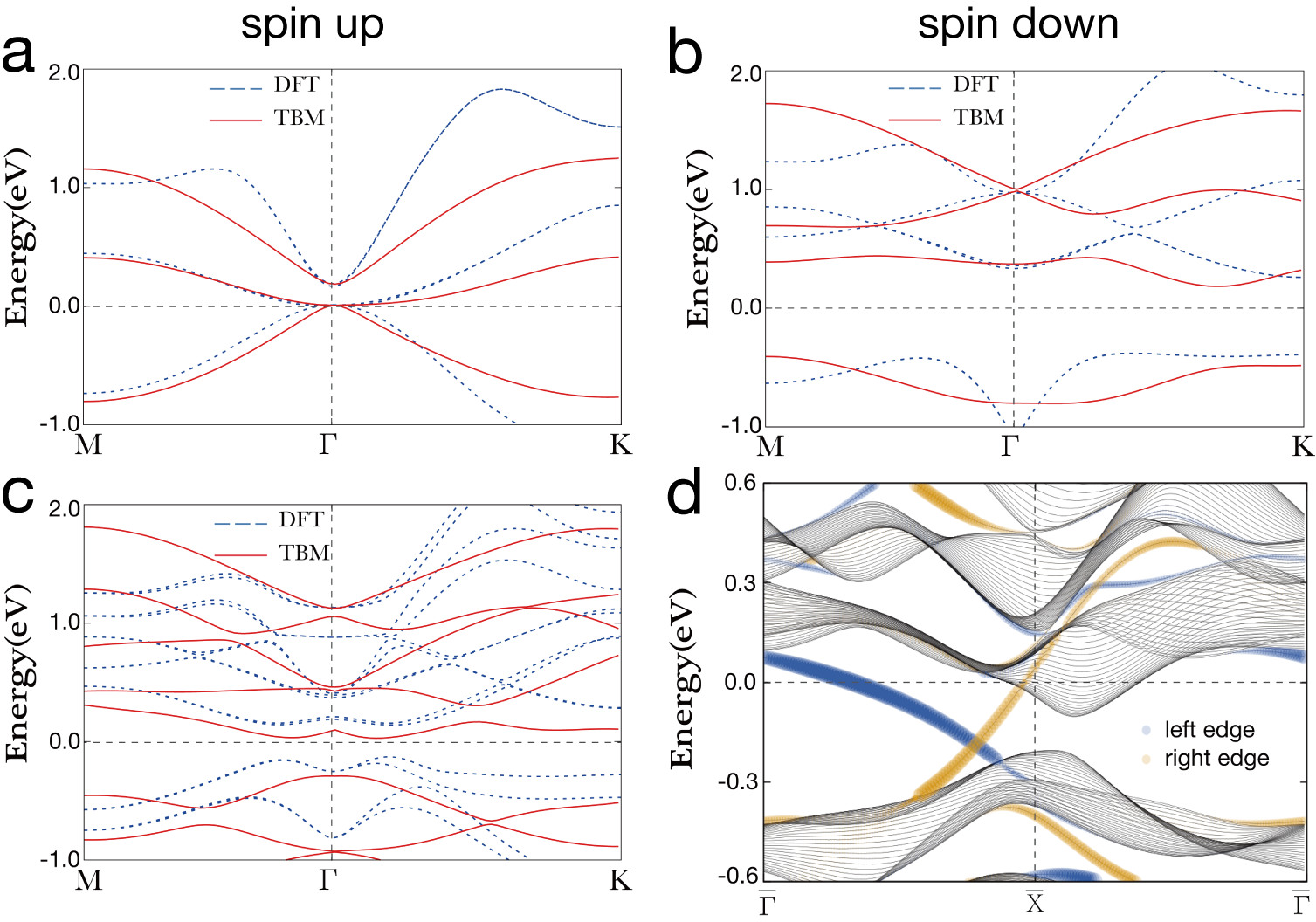}
\caption{(Color online) Comparison of the $p$-orbital model in Eq.~(\ref{Eq:model}) and the DFT calculations. In (a-c), the blue dashed lines and the solid red lines correspond to the DFT results and the tight-binding fitting, respectively. (a-b) Spin-polarized electronic structure without SOC. (c) Spin-polarized electronic structure with SOC.  (d) The topological edge states calculations using the tight-binding model in a ribbon geometry with a zigzag edge. The blue and yellow circles represent the probability of the wave functions residing at the left edge and the right edge.}
\label{Fig2}
\end{figure}

Because the presence of the intrinsic magnetization breaks time-reversal symmetry, we use two different sets of parameters to construct $H_{0}$. 
Discarding all terms breaking spin-conservation which are not important at the level of our discussion, the Hamiltonian $H_{0}$ ($12 \times 12$) can be cast as a direct sum of two $6 \times 6$ Hamiltonian for the two spin sectors:
\begin{equation}\label{H_0}
H_{0} = \left(\begin{array}{cc}
H_{\uparrow \uparrow} & 0   \\
0 & H_{\downarrow \downarrow}
\end{array}\right)
\end{equation}
with
\begin{equation}
H_{\sigma \sigma} = \left(\begin{array}{cccccc}
\epsilon_{Bi,px}^{\sigma} & 0  & 0 & h_{xx}^{\sigma} & h_{xy}^{\sigma} & h_{xz}^{\sigma}  \\
0 & \epsilon_{Bi,py}^{\sigma} & 0 &h_{yx}^{\sigma} & h_{yy}^{\sigma} & h_{yz}^{\sigma} \\
0 & 0 & \epsilon_{Bi,pz}^{\sigma} & h_{zx}^{\sigma} & h_{zy}^{\sigma} & h_{zz}^{\sigma} \\
\dagger & \dagger & \dagger & \epsilon_{In,px}^{\sigma} & 0 & 0  \\
\dagger & \dagger & \dagger & 0 & \epsilon_{In, py}^{\sigma} & 0       \\
\dagger & \dagger & \dagger & 0 & 0 & \epsilon_{In,pz}^{\sigma}
\end{array}\right)\;,
\end{equation}
where $\dagger$ represents complex conjugation. 
We found that, to decently reproduce the DFT electronic structure, it is sufficient to consider only the nearest neighbor hopping between different sites. 
More details of the model parameters can be found in the Supplementary Information. 

In N-BiIn/Al$_2$O$_3$, the direction of magnetic polarization $\hat{\mathbf{m}} = (0,0,1)$ is normal to the monolayer plane.
Consequently, the Zeeman term only contributes diagonal elements to the effective Hamiltonian, i.e., intrinsic magnetization only leads to the spin-dependent energy shift and does not change the shape of the bands. 
We denote the new onsite energy level as
\begin{equation}\label{onsite}
\Delta_{\alpha}^{\sigma} = \epsilon_{\alpha}^{\sigma} \mp \lambda_{m}^{\alpha}
\end{equation}
where $-$ and $+$ correspond to spin-up and spin-down electrons. 

As for the SOC, we will only consider the simplest atomic SOC contribution, while discarding the Rashba contribution.
The latter only affects the details of the electronic structure, but is not essential to the topological phase transition.  
\begin{subequations}\label{SOC}
\begin{align}
&\left\langle p_{y}|\vec{L} \cdot \vec{S}| p_{x}\right\rangle=i \sigma_{z} \\
&\left\langle p_{z}|\vec{L} \cdot \vec{S}| p_{x}\right\rangle=-i \sigma_{y} \\
&\left\langle p_{z}|\vec{L} \cdot \vec{S}| p_{y}\right\rangle=i \sigma_{x}
\end{align}
\end{subequations}
Due to the presence of $p_{z}$ orbital in this model, additional SOC terms between $p_{z}$ and $p_{x/y}$ orbitals appear as compared to the bismuthene model~\cite{Reis287, PhysRevB.98.165146}. 
Adding SOC to the tight-binding model opens a large global gap and induces a band inversion at $\Gamma$.

To obtain these model parameters, we mainly fit three spin-up bands around the Fermi level and three spin-down ones around 1 eV, as shown in Fig.~\ref{Fig2}(a-b). 
Other bands stay at higher binding energies. Thus, the quality of fitting on these bands is not crucial.
Here, we primarily fit the spin-up bands as it determines the topological nature and gap size of the system. On the other hand, the spin-down bands stay away from the Fermi level and are less critical to the low-energy model.   
In Fig.~\ref{Fig2}, DFT bands are shown as blue dashed lines. Bands from the fitted tight-binding model are shown as solid red lines. 
Spin-up and spin-down bands are different in shape, but degeneracy and symmetry remain identical. 
Using two different parameter sets for the spin-up and spin-down electrons will capture such a difference. 
When SOC is included, band inversion between the $p_x$ and $p_y$ orbitals occurs in the spin-up and spin-down bands, see appendix Fig.~S4 for all three systems. 
However, as the Zeeman field shifts the electronic structure of the spin-down electrons upwards by 1 eV, only band inversion in the spin-up sector remains at the Fermi level.
Consequently, we effectively have a spin-polarized electronic structure with low-energy physics dominated by spin-up electrons only. 
The topological gap, thus, stems only from one spin component. 
{\it As a result, the topological nature and the nontrivial edge mode depend only on one spin sector, i.e., a single helical edge mode is obtained. }
We confirm the above analysis by directly calculating the edge states in Fig.~\ref{Fig2} (d), where blue and yellow bands correspond to the edge modes residing at the two different terminations in a ribbon calculation. 
Along each edge, there is only one topological mode consistent with the DFT calculations shown in Fig.~\ref{Fig1} (h) and confirms the QAH nature of the proposed model in Eq.~(\ref{Eq:model}). 

The proposed model correctly captures the QAH nature of N-BiIn/Al$_{2}$O$_{3}$, and it is generic to all four material systems studied in this work. 
In supplementary Fig.~S7, we further show the comparison of the fitted model to the DFT calculations for other materials, as well as the calculated helical edge modes from the model. 
The first three rows correspond to the spin-up, spin-down, and SOC bands. 
Our tight-binding model can explain the topology of a large variety of material systems. 
Replacing Bi with other group-V elements, such as Sb and As, is expected to result in similar band structures and the same QAH states. 
Although Al$_{2}$O$_{3}$ may not be a good substrate candidate anymore in those cases due to lattice mismatch, the model we proposed most likely still applies. 

\section{Discussions and Conclusions}

\begin{figure}[htbp]
\centering
\includegraphics[width=\linewidth]{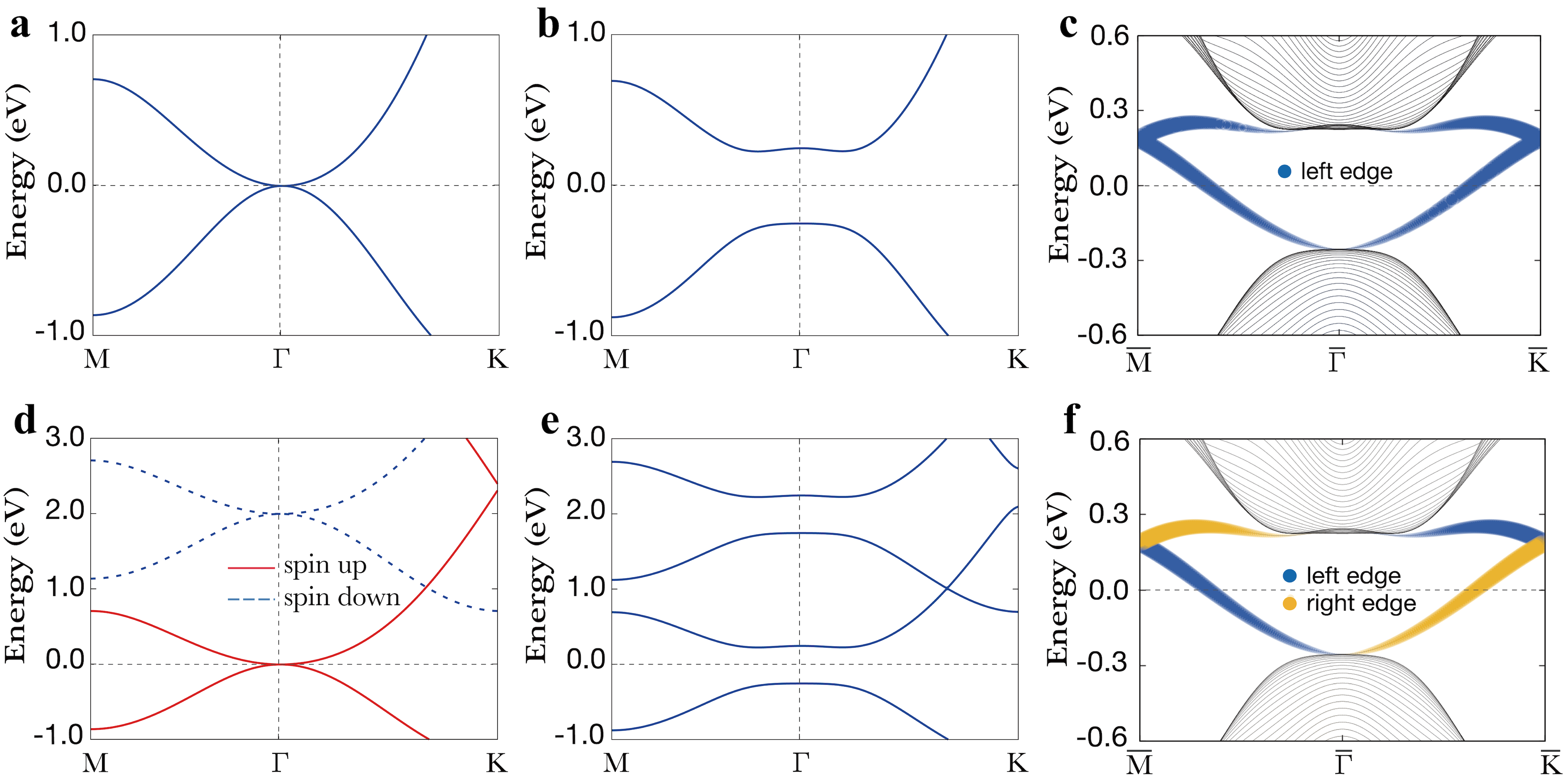}
\caption{(Color online) Electronic structure and topological phase transition in a $p_{x/y} + $ Zeeman model. (a -c) and (b - d) correspond to the 
TR symmetric and symmetry-broken cases. Electronic structures in (a, b, d, e) correspond to the calculations of the simplified model with (a) $H_{0}$, (b) $H_{0}+H_{SO}$, (d) $H_{0} + H_{M}$, and (e) $H_{0}+H_{M}+H_{SO}$ terms, respectively.
(c) and (f) show the topological edge modes for the QSH and QAH states, respectively.}
\label{Fig3}
\end{figure}

We have proposed using two different parameter sets to reproduce the DFT electronic structure. 
In this section, we further simplify the proposed model and explain the topological phase transition more transparently.

To obtain a universal effective modeling through simplification, we have identified two aspects of the full-scale generic model in Eq.~(\ref{Eq:model}) that turn out to be irrelevant for the topological nature and QSH-QAH phase transition in honeycomb-type electronic p-orbital models:
(i) the adoption of different parameter sets for spin-up and spin-down bands, and (ii) the $p_{z}$ orbitals. 
Discarding them from Eq.~(\ref{Eq:model}), we arrive at a model on the following basis. 
$\left|p_{x \uparrow}^{Bi}\right\rangle,\left|p_{y \uparrow}^{Bi}\right\rangle,\left|p_{x \uparrow}^{X}\right\rangle,\left|p_{y \uparrow}^{X}\right\rangle,\left|p_{x \downarrow}^{Bi}\right\rangle,\left|p_{y \downarrow}^{Bi}\right\rangle,\left|p_{x \downarrow}^{X}\right\rangle,\left|p_{y \downarrow}^{X}\right\rangle$. 
$H_{0}$ is an $8 \times 8$ matrix consisting of two $4 \times 4$ matrices for the two spin sectors. 
\begin{equation}
H_{\uparrow\uparrow} = H_{\downarrow\downarrow} = \left(\begin{array}{cccc}
\epsilon_{Bi,px} & 0   & h_{xx} & h_{xy} \\
0 & \epsilon_{Bi,py}  &h_{yx} & h_{yy} \\ \\
\dagger & \dagger  & \epsilon_{X,px} & 0   \\
\dagger & \dagger  & 0 & \epsilon_{X,py}    
\end{array}\right)
\end{equation}
The matrix elements can be found in Eq.~(S1) of the Supplementary Information.
Here, we have neglected the spin dependence of all matrix elements and require them to be identical in both spin sectors.
The Zeeman splitting and the SOC term also take the same form as in Eq.~(\ref{onsite}) and Eq.~(\ref{SOC}).

This simplified model is sufficient to explain the topological phase transition induced by nitrogen absorption.
We note that such a tight-binding model is generic to all two-dimensional spin-polarized $p_{xy}$-orbital system~\cite{Kaloni2014, PhysRevLett.113.256401, Chen2016, Huang2018, HUANG2020246}.
In Fig.~\ref{Fig3}, we show the electronic structure of the simplified model. 
Without SOC and Zeeman splitting, this model shows two parabolic bands touching at the Fermi level as displayed in Fig.~\ref{Fig3}a. 
Each band is doubly degenerate due to spin. SOC relieves the band degeneracy at $\Gamma$, leading to a topological QSH. 
This model contains two counter-propagating edge modes inside the bulk energy gap, as shown in Fig.~\ref{Fig3} (c). 
Half-passivation of nitrogen leads to the onset of a long-range ferromagnetic order, which induces the Zeeman splitting in our model. 
Upon turning on the Zeeman splitting, the spin-degenerate bands shown in Fig.~\ref{Fig3} (a) start to separate. 
%The spin-up and spin-down bands move downward and upwards in energy, respectively. 
We keep the spin-up band at the Fermi level by modifying the chemical potential, see Fig.~\ref{Fig3}d. 
The separation of the two bands does not modify the topology of either band. When SOC is further included in Fig.~\ref{Fig3} (e), both bands become gapped and attain a topological character. 
Because the Fermi level is now only governed by the spin-up bands, however, even though the spin-down bands at 2 eV are topological as well, this model will only demonstrate one helical edge mode on each edge, as shown in Fig.~\ref{Fig3} (f). 
Thus, in Bi-III monolayer/Al$_{2}$O$_{3}$, the topological phase transition between QSH and QAH is mainly induced by Zeeman splitting. Both spin-up and spin-down bands remain topological, but one of them is removed from the Fermi level by the Zeeman field. 
Consequently, the aspired QAH state is reached. 
Here, magnetic order does a trivial job, and it induces the topological phase transition only by shifting the topological gap of one spin sector away from the 
Fermi level, while not destroying its topology. The latter would require strong off-diagonal spin-exchange couplings, which are small in the materials we have considered.

In summary, through first-principle calculations, we propose that Al$_{2}$O$_{3}$ can be a promising substrate candidate to grow binary monolayers consisting of group III elements (Al, Ga, In, Tl) and bismuth (Bi). 
These systems belong to the same category of the large-gap QSH states as bismuthene/SiC(0001), with topological gaps of hundreds of meV. 
When further half-passivated with nitrogen, long rang ferromagnetic order is induced by breaking the complete shell configuration of N valence electrons. 
Consequently, a transformation from QSH to QAH phases occurs. The topological gap remains considerably large in the QAH state, which promises a great chance to be in reach for application in spintronic devices.  

We have further provided an analytical understanding of their low-energy topological physics by constructing a generic tight-binding model, which reveals that the topological phase transition between QSH and QAH is essentially induced by Zeeman splitting. 
The nontrivial character of both the spin-up and spin-down bands remain unaffected by the onset of ferromagnetism. 
At the same time, the topological gap of one spin is removed from the Fermi level, which effectively creates a spin-polarized half-metal and, thus, the QAH under SOC. 
Our work provides a generalization of the bismuthene platform to the case of broken time-reversal symmetry. We believe that the proposed tight-binding model applies to all similar systems dominated by $p_{x}, p_{y}$ orbitals, which will allow us to supplement the experimental effort with theoretical guidance along with the search for large-gap QAH systems.  

\section{Acknowledgement}
This work was supported by the National Key R$\&$D Program of China (2017YFE0131300), Sino-German mobility program (M-0006), National Natural Science Foundation of China under Grant No. 11874263, and Shanghai Technology Innovation Action Plan 2020-Integrated Circuit Technology Support Program (Project No. 20DZ1100605). W.S. wants to thank the financial support of Science and Technology Commission of Shanghai Municipality (STCSM) (Grant No. 22ZR1441800), Shanghai-XFEL Beamline Project (SBP) (31011505505885920161A2101001). 
This work in W\"urzburg is funded by the Deutsche
Forschungsgemeinschaft (DFG, German Research Foundation) through
Project-ID 258499086 - SFB 1170 and through the W\"urzburg-Dresden
Cluster of Excellence on Complexity and Topology in Quantum Matter - ct.qmat Project-ID 390858490 - EXC 2147.
Part of the calculations were performed at the HPC Platform of ShanghaiTech University Library and Information Services, at the School of Physical Science and Technology, and at the Scientific Data Analysis Platform of Center for Transformative Science.

\bibliographystyle{apsrev4-1}
\bibliography{ref}  

\end{document}